\documentclass[letter]{aa}  
\usepackage{natbib}
\usepackage{braket}
\usepackage{graphicx}
\usepackage{booktabs}
\usepackage{setspace}
\usepackage{comment}
\usepackage[colorlinks=true,linkcolor=blue,citecolor=blue, urlcolor=blue]{hyperref}
\usepackage{txfonts}

\usepackage{color}
\usepackage{breqn}
\usepackage{amsmath}
\usepackage[normalem]{ulem} % For \sout{}

\newcommand{\soutPC}{\bgroup\markoverwith{\textcolor{cyan}{\rule[0.5ex]{2pt}{1pt}}}\ULon}

\newcommand{\soutFS}{\bgroup\markoverwith{\textcolor{red}{\rule[0.5ex]{2pt}{1pt}}}\ULon}
\begin{document}

   \title{The baryonic Tully-Fisher relation as an independent direct\\ probe of cosmology and of the nature of dark matter}

   \titlerunning{The BTFR as a direct probe of cosmology and dark matter nature}

   \author{Francesco Sinigaglia      \inst{1,2,3,4}\fnmsep\thanks{IFPU Fellow; \email{fsinigag@sissa.it}}
          %\and
          %Matteo Viel\inst{1,2,3,4,5}
          }

   \authorrunning{F. Sinigaglia}

   \institute{Institute for Fundamental Physics of the Universe (IFPU), Via Beirut 2, I-34151 Trieste, Italy
        \and SISSA - International School for Advanced Studies, Via Bonomea 265, 34136 Trieste, Italy
        \and INAF - Osservatorio Astronomico di Trieste, Via G. B. Tiepolo 11, I-34131 Trieste, Italy
        \and INFN – National Institute for Nuclear Physics, Via Valerio 2, I-34127 Trieste, Italy
        %\and ICSC - Centro Nazionale di Ricerca in High Performance Computing, Big Data e Quantum Computing, Via Magnanelli 2, Bologna, Italy
             }

   \date{Received \today; accepted XYZ}

% \abstract{}{}{}{}{} 
% 5 {} token are mandatory
 
  \abstract
   {The baryonic Tully-Fisher relation (BTFR), a well-established galaxy scaling relation linking the dynamical mass of rotation-supported galaxies through their maximum circular velocity to the baryonic luminous mass, has emerged over the decades as a fundamental scaling relation and as a robust calibrated distance indicator, thereby providing a robust benchmark to test galaxy formation and evolution theories as well as an independent probe of the expansion of the Universe.}
   {In this Letter, we show for the first time that the BTFR is also simultaneously directly sensitive to the cosmological parameters $\Omega_m$ and $\sigma_8$, the astrophysical feedback from supernovae (SNe) and active galactic nuclei (AGN), and the mass of dark matter particles $M_{\rm wdm}$, and can therefore be used as novel, direct probe of cosmology and fundamental physics.}
   {We perform simulation-based inference on the large \texttt{DREAMS} cosmological magneto-hydrodynamic simulations suite and train deep neural networks in the form of normalizing flows to estimate the posterior distributions of $\Omega_m$, $\sigma_8$, $M_{\rm wdm}$ and the three astrophysical free parameters, given a BTFR measurement.}
   {Our framework is able to recover unbiased values for $\Omega_m$ and $\sigma_8$, with subpercent deviations accuracy and a $\sim 2.6\%$ and $\sim 3.9\%$ median precision, respectively, to capture the warm dark matter particle mass $M_{\rm wdm}$ within a $\sim 30-35\%$ precision, as well as to constrain the SN feedback parameters (but not the one regulating AGN feedback).}
   {We conclude that, beyond its usage as a distance indicator and to constrain the baryon cycle and the feedback mechanisms shaping galaxy formation and evolution, the BTFR constitutes a direct independent probe of cosmology and fundamental physics and opens new promising avenues, such as constraining the cosmic growth rate and testing gravity theories, to be explored with the future Square Kilometer Array.}

   \keywords{Cosmology: dark matter, dark energy --- Galaxies: formation, evolution --- Methods: statistical}

   \maketitle 

%-------------------------------------------------------------------
%-------------------------------------------------------------------
%-------------------------------------------------------------------

\section{Introduction}
\label{sec:intro}

The baryonic Tully–Fisher relation \citep[hereafter BTFR, e.g.,][]{McGaugh2000} is an empirical scaling law that links the total baryonic mass of a rotationally-supported galaxy to its asymptotic rotation velocity. As an extension of the original Tully–Fisher relation \citep{TullyFisher1977}, which connected luminosity to rotation velocity, the BTFR probes the gravitational potential of the host dark matter halo through its HI disk, reducing scatter and improving its universality across a wide range of galaxy masses and morphologies \citep[e.g.,][]{McGaugh2000,McGaugh2012,denHeijer2015,Lelli2016,Sales2017}. %Over the past two decades, increasingly precise measurements of galaxy kinematics and baryonic content have established the BTFR as a remarkably tight correlation, with important implications for galaxy formation theories and the distribution of dark matter.
Beyond its role in galaxy formation and evolution, the BTFR (as an extension of the more traditional TFR) has been regarded as a powerful calibrated distance indicator   \citep[e.g.,][]{Schombert2020,Kourkchi2022} alongside other methods, offering a complementary approach that is particularly well suited to gas-rich, late-type galaxies. In this sense, the BTFR can effectively be used to estimate $H_0$ and estimate the expansion rate of the Universe \citep[e.g.,][]{Schombert2020}.

Nevertheless, the BTFR has not yet been widely exploited as a tool for constraining cosmological parameters. Most applications to date have focused on testing galaxy evolution and dynamics \citep[e.g.,][]{McGaugh2012,Lelli2016,Sorce2016,Ponomareva2018,Ponomareva2021,Sharma2024,Zoonozi2025}, probing dark matter halo properties \citep[e.g.,][]{Trachternacht2009,2009A&A...507..635S,Kang2013}, or evaluating alternative theories of gravity \citep[e.g.,][]{McGaugh2012,McGaugh2016,Lelli2017,Borka2025}. The use of the BTFR in a cosmological context, apart from its usage as a distance indicator, has remained underdeveloped. If successful, using the BTFR as a cosmological probe would offer a novel, independent cosmological probe at low redshift, which would potentially help shed light on current tensions between different measurements of cosmological parameters \citep[see e.g.,][for recent reviews]{DiValentino2021,Giare2025,Pantos2026}. 

In this work, we use a simulation-based inference (SBI) framework relying on deep neural networks to explore for the first time whether the BTFR is sensitive to a broader range of cosmological parameters beyond $H_0$, as well as to the dark matter mass, thereby constituting an independent and competitive probe of cosmology and fundamental physics. %In particular, because both the total baryonic mass and the asymptotic rotation velocity depend directly on the halo mass and other halo properties, which is turn has a direct dependence on cosmology, one intuitively expects the BTFR to be sensitive on the underlying cosm. 

The paper is organized as follows. Section~\ref{sec:dreams} introduces the \texttt{DREAMS} simulations. We summarize the simulation-based inference framework adopted in this work in Section~\ref{sec:sbi}, and present  results and discussion in Section~\ref{sec:results}. We conclude in Section~\ref{sec:conclusions}.

%-------------------------------------------------------------------
%-------------------------------------------------------------------
%-------------------------------------------------------------------

\section{The BTFR from the \texttt{DREAMS} simulations}\label{sec:dreams}

The {\it DaRk mattEr and Astrophysics with Machine learning and Simulations} project \citep[hereafter \texttt{DREAMS},][]{Rose2025} comprises a large suite of state-of-the-art cosmological N-body and magneto-hydrodynamic simulations, aimed at exploring the joint interplay of cosmology, astrophysics and the nature of dark matter in shaping galaxy formation. The \texttt{DREAMS} simulations comprise cosmological boxes, as well as Milky-way and dwarfs zooms-in; we focus in this paper on the former. Similarly to the related subset of the `parent' \texttt{CAMELS} project \citep{VillaescusaNavarro2021}, the \texttt{DREAMS} boxes were run using the \texttt{AREPO} code \citep{Springel2010,Weinberger2020} and the \texttt{IllustrisTNG} model \citep{Weinberger2017,Pillepich2018,Nelson2019} in cosmological volumes $V=(25~h^{-1}~{\rm Mpc})^3$ with periodic boundary conditions, following the evolution of $256^3$ dark matter particles of mass $m_{\rm dm}=6.44\times 10^7(\Omega_m-\Omega_b)/0.251~h^{-1}~{\rm M_\odot}$ and of $256^3$ gas resolution elements with an initial mass $m_{\rm gas}=1.27 \times 10^7~h^{-1}~{\rm M_\odot}$. The initial conditions were generated at $z=127$ using second-order Lagrangian Perturbation Theory and the cosmological parameters (except for $\Omega_m$ and $\sigma_8$) were fixed to the following values: $\Omega_b=0.049$, $h=6771$, $n_s=0.9624$, $M_\nu=0.0~{\rm eV}$, $w=-1$, and $\Omega_k=0$. The $1,024$ simulations designed to sample $6$ different parameters --- the two cosmological parameters $\Omega_m$ and $\sigma_8$, the three astrophysical parameters parametrizing supernovae and AGN $A_{\rm SN1}$, $A_{\rm SN21}$, $A_{\rm AGN}$, and the mass of the warm dark matter particles $M_{\rm WDM}$ --- following a Sobol sequence in the following ranges: $\Omega_m\in[0.1,0.5]$, $\sigma_8\in[0.6,1.0]$, $A_{\rm SN1}\in[0.25,4.0]$, $A_{\rm SN2}\in[0.5,2.0]$, $A_{\rm AGN1}\in[0.25,4.0]$, $M_{\rm WDM}\in[1.8,16]$ keV. We notice that the heaviest dark matter particles probed by the suite produce an initial matter power spectrum which is indistiguishable from the Cold Dark Matter (CDM) one up to the Nyquist frequency $k=32~h~{\rm Mpc}^{-1}$, and are therefore equivalent to standard CDM simulations \citep{Rose2025}. 

To derive the BTFR for each simulation we proceed as follows. The total baryon mass of each galaxy is computed as the sum of the stellar and the gas particles: $M_{\rm b}=M_\star + M_{\rm gas}$. As a proxy of the maximum circular velocity for the baryonic component $V_{\rm max,b}$, we assume the maximum circular velocity of the host subhalo, considering all the particle species including dark matter. While in principle this $V_{\rm max}$ is strictly speaking not the same as $V_{\rm max,b}$, they can be approximated to be the same for Milky-way-type galaxies \citep[e.g.,][]{Dutton2010}. In practice, we derive this information from the \texttt{SUBFIND} \citep{Springel2001,Dolag2009} catalog  (i.e., selecting the \texttt{Subhalo} key) made publicly available. Since the BTFR applies only for rotationally-supported galaxies, we select only galaxies whose ratio $V_{\rm max}/\sigma_{\rm disp}>1.5$, where $\sigma_{\rm disp}$ is the velocity dispersion. We explicitly verified that enforcing more conservative cuts %--- $V_{\rm max}/\sigma_{\rm disp}>2$ and $V_{\rm max}/\sigma_{\rm disp}>3$ --- 
does not improve the results but just reduce the available sample. In addition, we enforce a cut in the maximum circular velocity $\log_{10}(V_{\rm max}/({\rm km~s^{-1}}))>1.5$, to avoid major resolution issues. We discuss the potential resolution effect stemming from the dependence of the dark matter particle mass on the cosmology in Appendix~\ref{app:particle_mass}. After applying these cuts, we perform a random downsampling to ensure that the final sample has the same size $n=1000$ in every $\texttt{DREAMS}$ simulation. This is done to prevent the training of the neural network to be sensitive to the inhomogeneous sample size across realizations, and hence, to be potentially biased to some specific ones. For each \texttt{DREAMS} realization, we represent the BTFR by using $10$ bins in $V_{\rm max}$, and then computing the median baryonic mass $M_{\rm b,med}$ and its standard deviation $\sigma_M$. We extract the BTFR at three redshift: $z=0$, $z=0.54$, and $z=1.05$, and combine them altogether in the main analysis. We target this redshift range, as it will be the one at which the BTFR will be reliably probed by the SKA. We show examples of the derived BTFR in Figure~\ref{fig:btfr_pars} in Appendix~\ref{app:bftr_fix} and discuss the gain in performing a multi-redshift versus single-redshift analysis in Appendix~\ref{app:single_multi_z}.

%-------------------------------------------------------------------
%-------------------------------------------------------------------
%-------------------------------------------------------------------

\section{The inference framework}\label{sec:sbi}

In this work, we aim at estimating the posterior distribution of the free simulation parameters given a BTFR measurement by applying SBI. Briefly, we adopt a neural posterior estimation (hereafter NPE) approach to directly emulate the posterior distributions through neural density estimation technique \citep[e.g.,][]{Paramakarios2019}. Specifically, we employ Masked Autoregressive Flows \citep[hereafter MAF,][]{Tabak2010,Tabak2013,JimenesRezende2015,Papamakarios2017}, relying on the implementation from the \texttt{sbi} package \citep{TejeroCantero2020,BoeltsDeistler_2025} through the \texttt{LtU-ILI} interface \citep{Ho2024}. We use as fiducial setup MAF with $10$ transforms, $100$ hidden units, and a batch size $n_b=64$. We adopt the \texttt{Adam} optimizer \citep{Kingma2015} with a learning rate $\eta=5\times 10^{-4}$. Out of the full set of $1,024$ simulations, we use $974$ as training set, and the remaining $50$ for testing, unless stated otherwise. The training set is split into $90\%$/$10\%$ training/cross-validation. To prevent overfitting, the training is stopped if no improvement is found after $20$ consecutive epochs. We have explicitly tested other network configurations with more transforms and/or hidden units, and found no improvement in the results. Per each simulation, we feed as input a one-dimensional data vector obtained by stacking $V_{\rm max}$, $M_{\rm b,med}$ and $\sigma_M$ arrays per each realization and redshift snapshot as input, and as output the true model parameters. %All the runs were performed using CPUs on an ordinary MacBook Pro with a 10-core 16GB M2 processor. The training took $<1$ minutes. The testing was performed by drawing $20,000$ posterior samples per each parameter and discarding the first $1,000$ to avoid the burn-in phase and took $<2$ minute per simulation, amounting to $\sim 1.5$ hours for the full sample. 
We describe more in detail our inference framework in  Appendix~\ref{app:workflow} and address the robustness of the posteriors calibration in Appendix~\ref{app:posterior_coverage}.

%-------------------------------------------------------------------
%-------------------------------------------------------------------
%-------------------------------------------------------------------

\begin{figure*}
    \centering
    \includegraphics[width=0.8\linewidth]{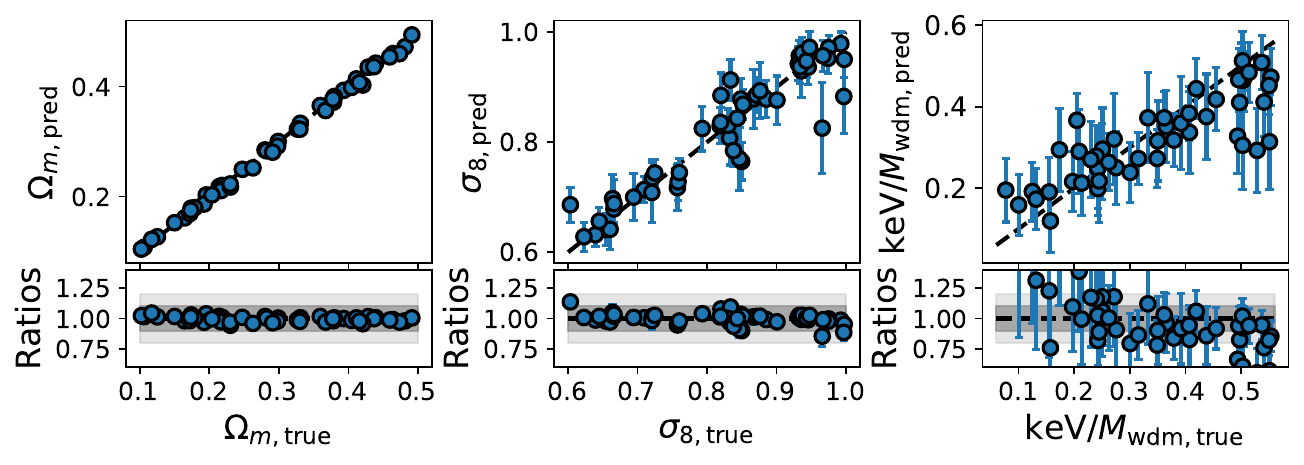}
    \includegraphics[width=0.8\linewidth]{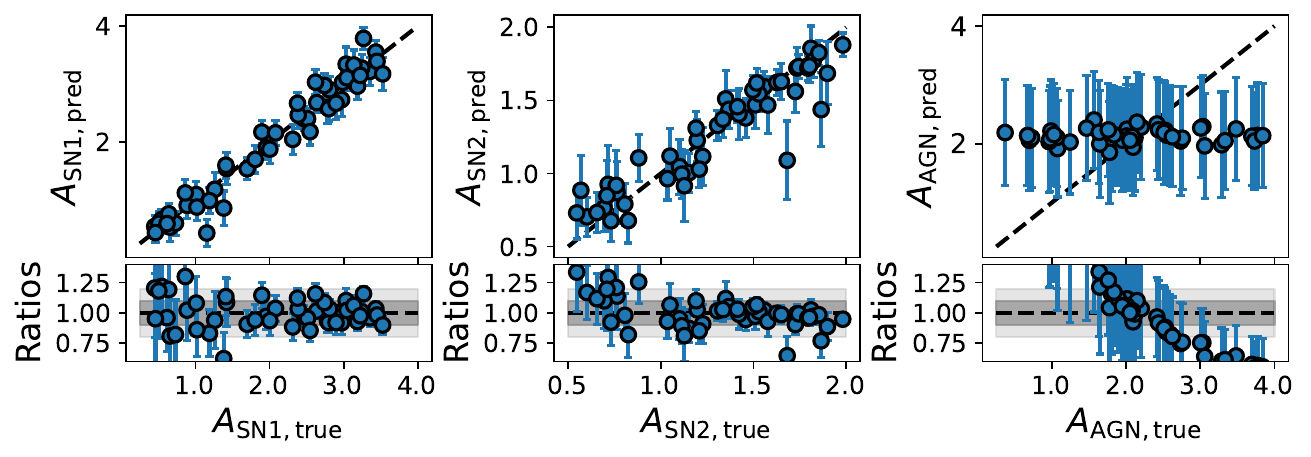}
    \caption{Predicted values for $\Omega_m$ (top left), $\sigma_8$ (top mid), ${\rm keV/M_{\rm wdm}}$ (top right), $A_{\rm SN1}$ (bottom left), $A_{\rm SN2}$ (bottom mid), and $A_{\rm AGN}$ (bottom right) against the true parameters values. The top subpanels show the 1:1 comparison, while the bottom subpanels display the ratios between the predicted and the true values, and the gray shaded regions stand for $10\%$ (darker) an $20\%$ (lighter) deviations.}
    \label{fig:residuals}
\end{figure*}

\section{Results and discussion}\label{sec:results}

To diagnose the results from our inference framework, we first inspect the predicted values for the parameters against the true ones from the simulations. The results are shown in Figure~\ref{fig:residuals}, for $\Omega_m$ (top left), $\sigma_8$ (top mid), and ${\rm keV}/M_{\rm wdm}$ (top right), $A_{\rm SN1}$ (bottom left), $A_{\rm SN2}$ (bottom mid), $A_{\rm AGN}$ (bottom right). All the parameters can be recovered, at least to some extent, except for $A_{\rm AGN}$ as an effect of the large shot noise --- arguably larger than the signal --- associated with sparseness of AGNs in a small cosmological box, as already discussed in other works \citep[e.g.,][]{VillaescusaNavarro2021}. These findings are consistent with the ones from the halo-to-halo variation analyses performed on the \texttt{DREAMS} zoom-in simulations \citep{Garcia2026,Leisher2026,Lilie2026}, which report that $A_{\rm AGN}$ has a lesser impact on the halo dark matter density profiles, velocity distributions, and merger trees. In particular, $\Omega_m$ and $\sigma_8$ are recovered with outstanding accuracy --- $\sim0.4\%$ and $\sim0.2\%$ average deviation across the $50$ realizations --- and precision --- $\sim2.1\%$ and $\sim4.8\%$ standard deviation deviation across the $50$ realizations; median precision $\sim 2.6\%$ and $\sim3.9\%$, respectively --- and the predictions feature a final Spearman rank-order correlation $R=0.99$ and $0.93$ with the true value, respectively. The WDM mass is recovered to a lesser extent ($R=0.83$), but the inference framework is still able to achieve an accuracy of $\sim 3\%$ average deviation, a precision $\sim 33\%$ in the standard deviation across the $50$ realization, as well as a $\sim 29\%$ median precision. Interestingly, as anticipated, our framework is sensitive to both the $A_{\rm SN1}$ ($R=0.98$) and the $A_{\rm SN2}$ ($R=0.93$) parameters, recovered with an accuracy of $1.0\%$ and $1.6\%$  average deviation, a precision of $\sim 15.1\%$ and $14.9\%$ standard deviation across the $50$ realizations, as well as a $10.5\%$ and $12.6\%$ median precision, respectively.

To develop a better intuition of the parameter covariance, we compute the correlation coefficient for the posterior samples for each realization, and then compute the summary statistics over the $50$ realizations. The $\Omega_m$ and $\sigma_8$ parameters are found to feature an anti-correlation ($R=-0.74$), typically found also from other probes. Interestingly, $\sigma_8$ and $M_{\rm wdm}$ anti-correlate too ($R=-0.40$).

The dependency of the BTFR on cosmology and on the mass of dark matter particles can be explained as follows. $\Omega_m$ and $\sigma_8$ have a similar `observational' (but not physical) effect on galaxy formation, which is also the reason why they are degenerate and anti-correlate: increasing the value of such parameters at fixed halo virial mass leads to earlier halo collapse, larger halo concentration, deeper gravitational potential, higher maximum circular velocity, making the galaxies rotate faster and also attracting more baryon mass within the gravitational potential. In contrast, the warmer the dark matter particles, the more the small-scale fluctuations are smoothed out in the initial conditions, the more the low-mass halo formation is suppressed and in general the later the haloes tend to form, and the lower the halo concentration. In this sense, $\Omega_m/\sigma_8$ and $M_{\rm wdm}$ have competing effects. We notice that $\Omega_m$ and $\sigma_8$, despite playing a similar role from an observational point of view, have a different cosmological meaning and effect: $\Omega_m$ represents the amount of gravitating matter and regulates the growth rate of cosmic structures, the expansion history, and the collapse times, while $\sigma_8$ sets the amplitude of primordial perturbations. As such, they have different dependencies with redshifts: $\Omega_m(z)=\rho_m(z)/\rho_c\propto(1+z)^3$, $\sigma_8\approx \sigma_{8,0} D(z)\propto 1/(1+z)$. For this reason, we argue that a multi-redshift analysis of the BTFR may help capturing this different dependency on redshift of the two cosmological parameters and break their degeneracy, as shown in Figure~\ref{fig:posteriors} in Appendix~\ref{app:single_multi_z}.
%, consistently with the fact that the warmer the dark matter (i.e., the smaller the dark matter particle) mass, the more the formation of lower-mass haloes is suppressed. Therefore, to be able to restore those haloes and fit the BTFR, a larger value of $\sigma_8$ is required.

The findings reported above have several important implications. For the first time, the BTFR is herein shown to be a direct probe of cosmological parameters, beyond its cosmological usage as a distance indicator. This opens up the possibility of using it as a new probe of cosmology at low redshift, completely independent from galaxy clustering, lensing, CMB, or other more traditional indicators. This fact is especially important and promising in the view of the SKA, which will observe the BTFR directly up to $z\sim 1$. The sensitivity of the BTFR on cosmology can potentially be leveraged to constrain cosmological parameters and to shed light on the $\sigma_8$ tension.  
Interestingly, we have also shown that the BTFR is sensitive to the dark matter particle mass. %, beyond astrophysics and cosmology. 
This offers the possibility to constrain dark matter models and fundamental physics. Future \texttt{DREAMS} simulations suites will include other models beyond CDM/WDM, such as the $h_{\rm peak}-k_{\rm peak}$ model, the Effective THeory of Structure formation \citep[ETHOS,][]{CyrRacine2016,Vogelsberger2016}, and atomic dark matter \citep{Roy2023}. 

While this study provides a theoretical proof of concept, in future work we will need to address the well-known issue of the lack of convergence in the predictions of different galaxy formation models, and the consequent difficulty in reliably perform cosmological inference on real data. %We discuss preliminary strategies in Appendix~\ref{app:hybrid}. %One possibility that we provide herein, and that will be thoroughly tested in future work, consists in using a hybrid definition of the BTFR, computing 
%$V_{\rm max}$ from the HI disk, but using only the stellar mass as baryon component n $M_{\rm b}$. This has several advantages: (i) the stellar mass predictions are generally better converged in numerical simulations than the ones for the gas mass, since $M_{\rm star}$ is more stable against instantaneous changes, while the gas reacts on a much shorter timescale, (ii) the $M_\star-M_{\rm halo}$ has been thoroughly studied in the literature and is generally robust against abundance matching methods. In particular, since the cosmological information is primarily driven by the Tully-Fisher relation defined by the dark matter haloes (i.e., $M_{\rm halo}$ vs $V_{\rm halo}$ from the halo), leveraging (ii) one can convert the stellar mass to halo mass and perform inference on the resulting hybrid Tully-Fisher relation. 

In addition, future studies adopting a SBI approach to the BTFR as a cosmological tool as done in this paper should carefully address the issue of modelling observational systematics, which we discuss in Appendix~\ref{app:obs}.

\begin{figure*}
    \centering
    \includegraphics[width=0.75\linewidth]{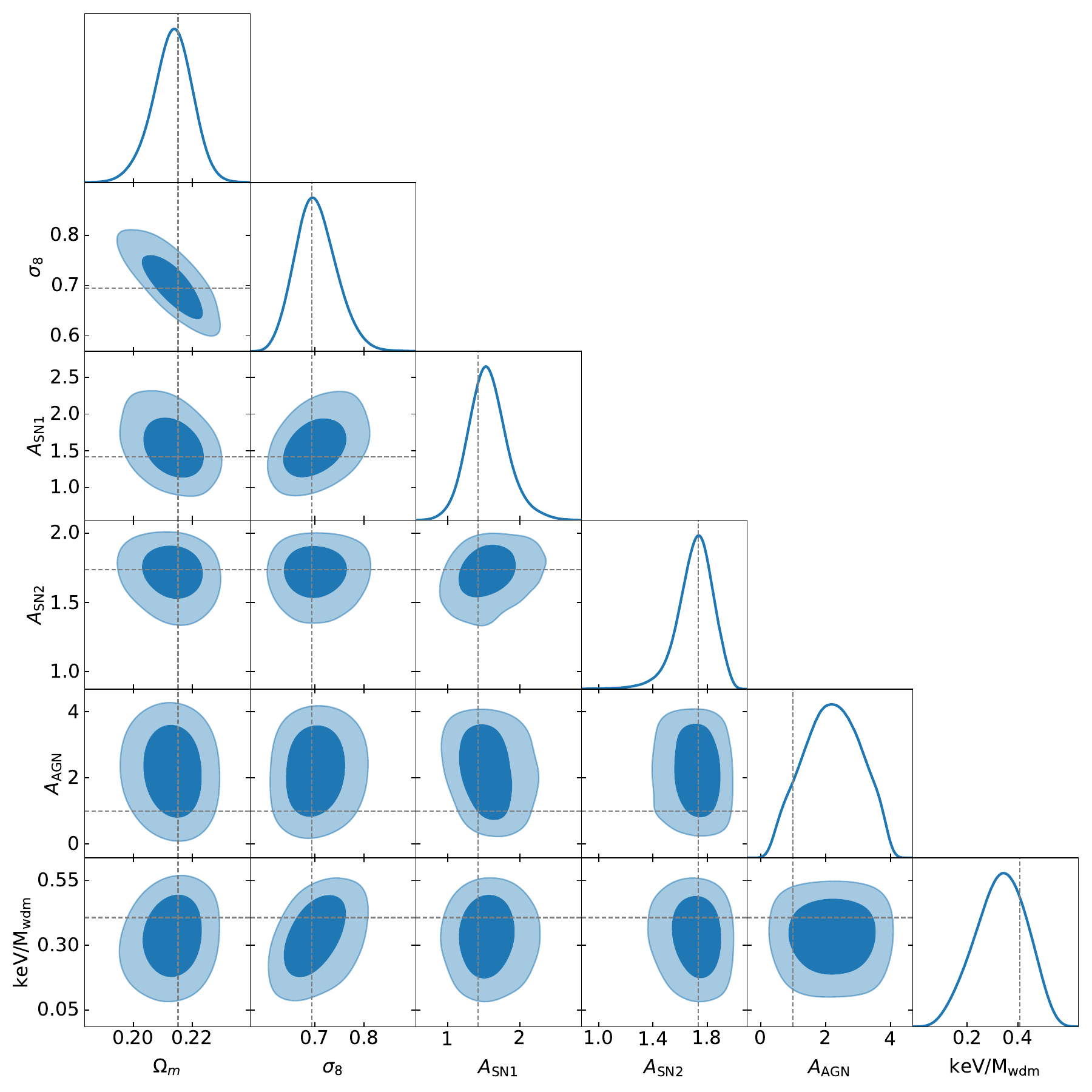}
    \caption{Posterior distributions of the model parameters for one random testing realization. The posteriors are shown as blue contours, while the gray dashed lines stand for the true parameter values.}
    \label{fig:posteriors_allpars}
\end{figure*}

%-------------------------------------------------------------------
%-------------------------------------------------------------------
%-------------------------------------------------------------------

\section{Summary and conclusions}\label{sec:conclusions}

In this work, we have presented a novel interpretation of the well-established baryonic Tully-Fisher relation \citep[BTFR]{McGaugh2000,McGaugh2012} as a direct independent probe of astrophysics, cosmology, and fundamental physics. In particular, while it is well-established that the BTFR depends on baryon feedback and the resulting baryon cycle driving the baryonic mass $M_{\rm b}$ in galaxies, we here show that we can directly infer the values of $\Omega_m$, $\sigma_8$ and the mass of WDM particles directly from the BTFR. By applying a simulation-based inference approach to the \texttt{DREAMS} cosmological hydrodynamic simulations, we show that we are able to correctly estimate $\Omega_m$ and $\sigma_8$ cosmological parameter with subpercent deviations accuracy and $\sim 2.6\%$ and $\sim 3.9\%$ median precision, respectively, as well as the WDM mass with $\sim 30-35\%$ precision (although with a potential systematics coming from a lower degree of correlation). Furthermore, the presented framework is able to correctly recover the values of the SNe feedback parameters, thereby offering the opportunity to precisely constrain feedback mechanisms. %The parameter describing AGN feedback is instead left unconstrained, due to the small cosmological volume available and the high shot noise associated with the sparseness of supermassive black holes. 

These findings open up new interesting avenues. In the first place, the BTFR is found to yield competitive cosmological constraints while being completely independent from other traditional cosmological probes --- such as galaxy clustering, weak lensing, the Lyman-$\alpha$ forest, and CMB observations, among others --- thereby offering an effective way of mitigating systematics. In particular, the BTFR offers a new low-redshift cosmological probe that can be used to further investigate e.g. the $\sigma_8$ tension. In addition, constraining the $\Omega_m$ and $\sigma_8$ parameters and their joint posterior distribution with the BTFR means estimating the $S_8$ parameter. This, together with the usage of the BTFR as a calibrator of peculiar velocities and for the consequent velocity power spectrum and $f\sigma_8$ estimation, allows to break degeneracies and isolate the growth rate $f$. This implies that a BTFR and peculiar velocity survey alone constitute a powerful probe of the cosmic growth rate and a test of gravity theories, without the need of external probes such as galaxy clustering or weak lensing. We will explore this idea in a forthcoming publication. In the second place, the Square Kilometer Array will measure the BTFR up to $z\sim 1$, which will allow us to estimate cosmological parameters across a large redshift range, thereby optimizing the extraction of information on the model parameters. We discuss the detectability of the BTFR from the SKA and its precursors in Appendix~\ref{app:ska_detectability}. In addition, combining the usage of BTFR as WDM probe to other constraints based e.g. on the Lyman-$\alpha$ forest \citep[see e.g.,][]{Irsic2024} provides a promising way forward to constrain the nature of dark matter. %Finally, this work investigates for the first time the covariance of the cosmological  of the dark matter mass on the BTFR, highlighting an anti-correlation between $\sigma_8$ and $M_{\rm wdm}$. %This advocates for the need of a comprehensive framework such as the ones presented herein. 

In forthcoming work, we will address the impact of the used galaxy formation model, along the lines discussed in this work, to guarantee a high degree of robustness when fitting real data, as well as the impact of observational effects on the BTFR model.

In conclusion, this work paves the way towards the full exploitation of the BTFR to estimate cosmological parameters, as well as constrain astrophysical processes shaping galaxy formation and evolution and shedding light on the nature of dark matter, on the way to the SKAO.

%-------------------------------------------------------------------
%-------------------------------------------------------------------
%-------------------------------------------------------------------

\begin{acknowledgements}
      F.S. is grateful to Matteo Viel, Sandeep Haridasu, Patricia Iglesias-Navarro, and Francisco-Shu Kitaura for useful discussions. F.S. acknowledges support from the {\it Institute for Fundamental Physics of the Universe} postdoctoral fellowship scheme. F.S. is grateful to Jonah Rose, Paul Torrey, Francisco Villaescusa-Navarro, Mariangela Lisanti and the whole \texttt{DREAMS} collaboration for making the simulations publicly available.%M.V. is partly supported by INFN INDARK and SISSA IDEAS grants, and by the INAF Theory Grant "Cosmological investigation of the cosmic web". M.V. acknowledges the funding by the European Union - NextGenerationEU, in the framework of the HPC project – “National Centre for HPC, Big Data and Quantum Computing” (PNRR - M4C2 - I1.4 - CN00000013 – CUP J33C22001170001).
\end{acknowledgements}

%\break 
%\clearpage

\bibliographystyle{aa}
\bibliography{lit}

\break 
\clearpage

\appendix

% ********************
\section{Dependence of the dark matter particle mass on the cosmological parameters}\label{app:particle_mass}

As noted in Section~\ref{sec:dreams}, the dark matter particle resolution $m_{\rm dm}\propto h^{-1}(\Omega_m-\Omega_b)$, which accounts for the varying mean density for different cosmologies, at fixed number of particles per simulation. While $\Omega_b$ and $h$ are kept fixed across the different \texttt{DREAMS} realizations, $\Omega_m$ is varied within a broad prior: $\Omega_m\in[0.1,0.5]$. This aspect introduces an indirect cosmology-dependent resolution effect, as the minimum resolved halo and galaxy masses depends on $\Omega_m$ at fixed number of particles. The cut in maximum circular velocity $\log_{10}(V_{\rm max}/({\rm km~s^{-1}}))>1.5$ partly alleviates this as it automatically excludes very low-mass galaxies, but it is not guaranteed that it lifts it completely. Furthermore, additional small resolution effects in the predictions of the galaxy baryon properties near the resolution limit may still be present. We will investigate this point in future work.

% *********************

\section{The BTFR at fixed parameters}\label{app:bftr_fix}

\begin{figure*}
    \centering
    \includegraphics[width=0.85\textwidth]{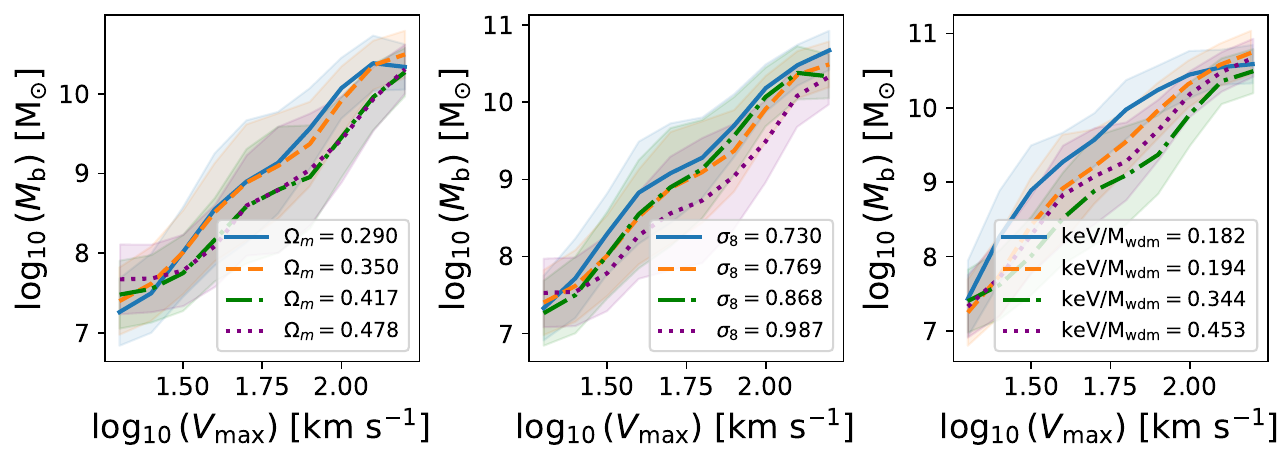}
    \includegraphics[width=0.85\textwidth]{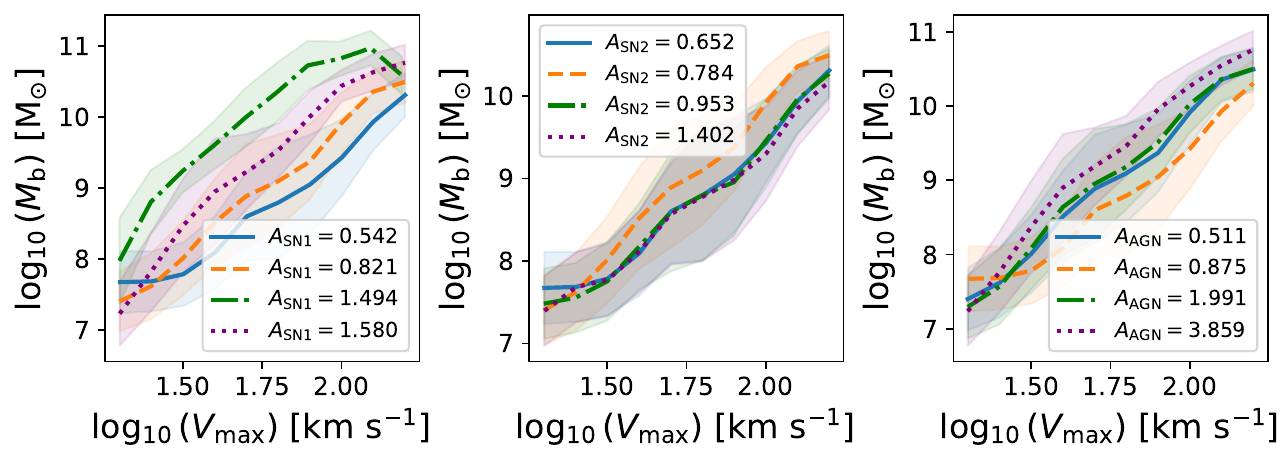}
    \caption{BTFR as extracted from different regions of the parameter space, by varying only one parameter at a time and keeping the others fixed (within a given narrow interval) : $\Omega_m$ (top left), $\sigma_8$ (top mid), ${\rm keV}/M_{\rm  wdm}$ (top right), $A_{\rm SN1}$ (bottom left), $A_{\rm SN2}$ (bottom mid), and $A_{\rm AGn}$ (bottom right).}
    \label{fig:btfr_pars}
\end{figure*}

To develop an intuitive understanding on the sensitivity of the parameters of the BTFR as extracted from the different \texttt{DREAMS} realizations, we show in Figure~\ref{fig:btfr_pars} the predictions for the BTFR from different regions of the parameter space, obtained by varying only one parameters and keeping the others fixed (within a given narrow interval). The parameter dependence affects all the three main parameters determining the BFTR: the slope, the normalization, and the scatter. It is hard to identify clear correlation just by a visual inspection. This suggests that the topology of the mapping between simulation and BFTR parameters is nontrivial, and support the usage of deep learning to unveil these dependencies.

% ********************
\section{Single vs multi redshift analysis}\label{app:single_multi_z}

\begin{figure*}
    \centering
    \includegraphics[width=0.5\linewidth]{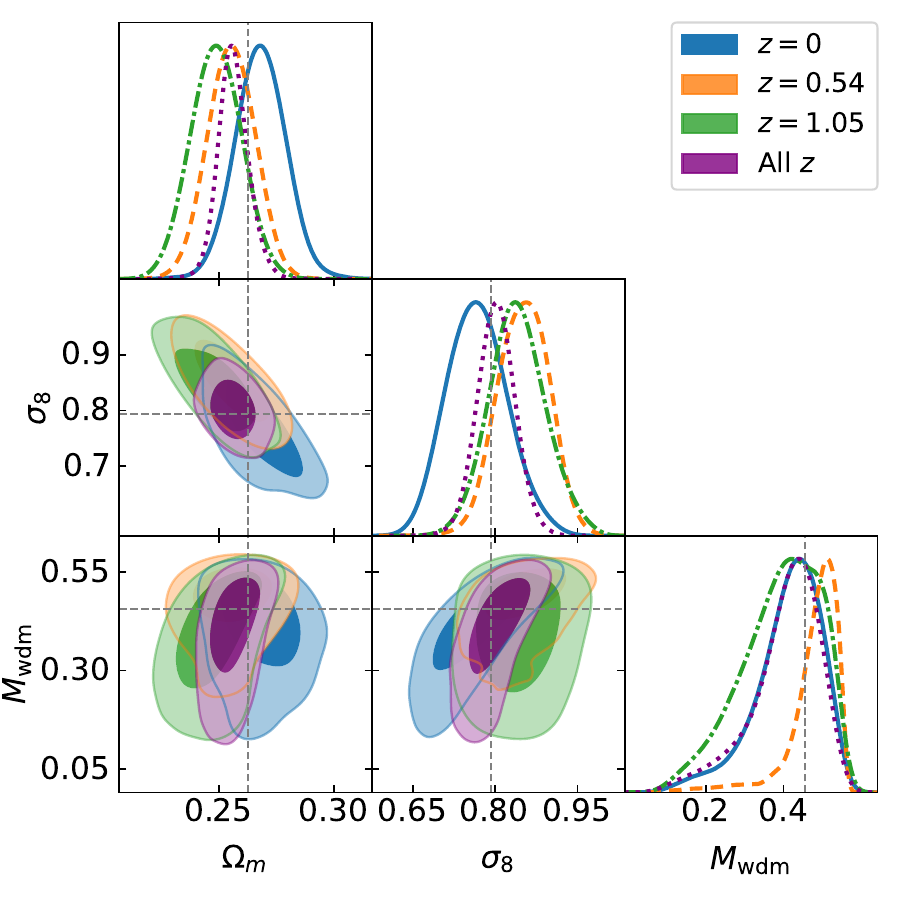}
    \caption{Posterior distributions of the $\Omega_m$, $\sigma_m$, and ${\rm keV}/M_{\rm wdm}$ parameters for one random testing realization, for different redshift selections: $z=0$ (blue), $z=0.54$ (orange), $z=1.05$ (green), the combination of the three (purple). The posteriors are shown as colored contours, while the gray dashed lines stand for the true parameter values.}
    \label{fig:posteriors}
\end{figure*}

To quantify the information gain in the cosmological parameters and the WDM mass stemming from combining different redshifts, as opposed to the analysis at a single redshift, we display the posterior distributions for one random testing realization in Figure~\ref{fig:posteriors}, separately for $z=0$ (blue), $z=0.54$ (orange), $z=1.05$ (green), and for all redshifts together (purple). A visual inspection reveals that the multi-redshift analysis improves significantly the accuracy and the precision of the results with respect to the single-redshift cases. We have explicitly verified that increasing the number of redshifts does not leave the results unchanged. %On the other hand, the improvement in the results is expected from the fact that $\Omega_m$ and $\sigma_8$ scale differently with redshifts (DETAILS HERE). 
Therefore, it is more important to consider a wider redshift range. We have herein tested the maximum redshift range $0<z<1$, which is the one that will be reliably probed by the SKA. 

% *********************

\section{Inference flowchart}\label{app:workflow}

\begin{figure*}
    \centering
    \includegraphics[width=0.85\linewidth]{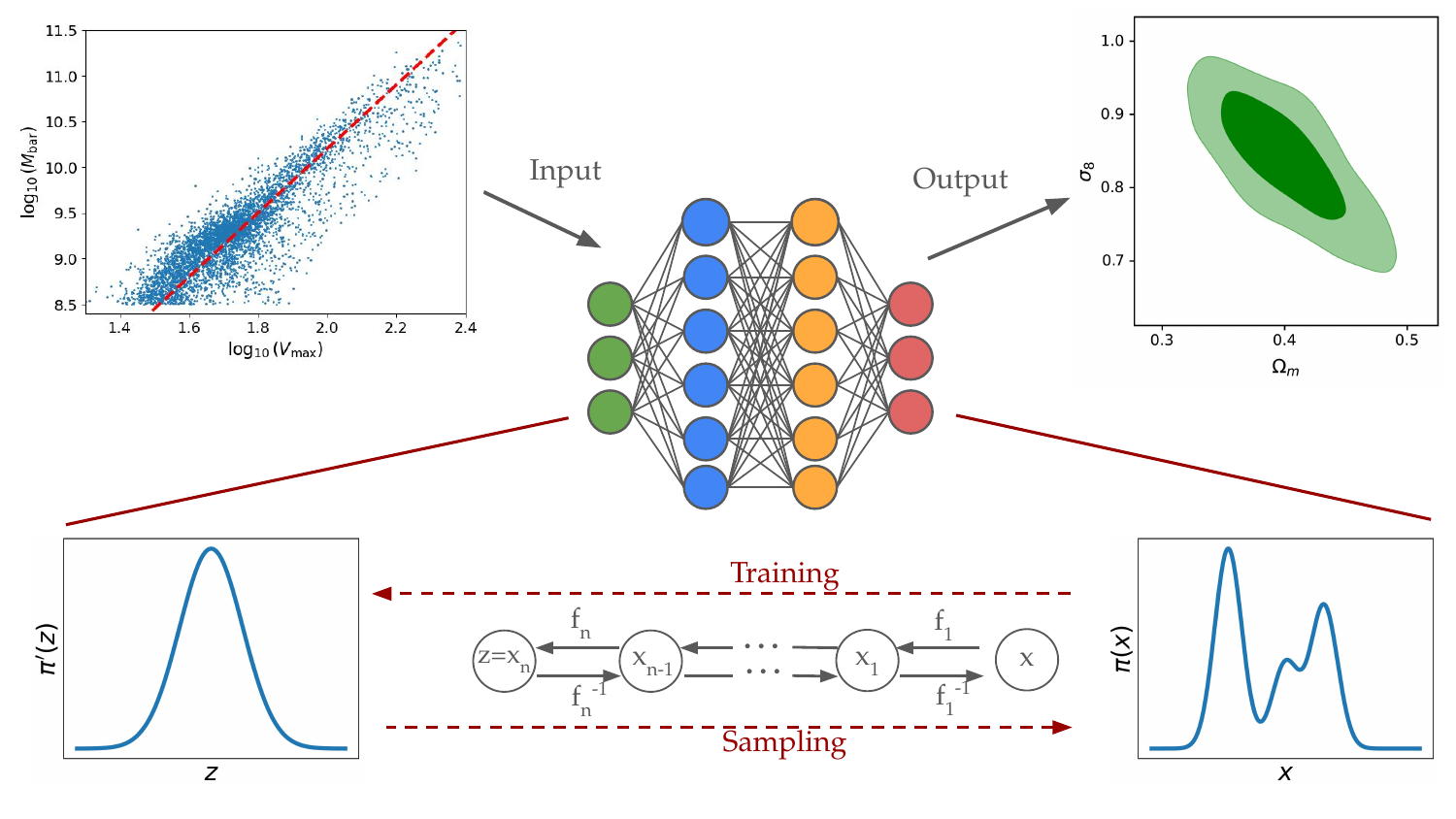}
    \caption{Graphical representation of the full workflow underlying this work. Top: extraction of the input BTFR (left) and example output posterior distributions. Mid: input/output setup of the neural network. Bottom: schematic representation of the normalizing flow.}
    \label{fig:workflow}
\end{figure*}

Figure~\ref{fig:workflow} provides a graphical representation of the full workflow. As already anticipated in Section~\ref{sec:sbi}, we first extract a measurement of the BTFR from each \texttt{DREAMS} realization (top left). Afterwards, we feed the training set to the normalizing flow and train it to retrieve the parameter posterior distributions (top right). In this phase (red dashed arrow pointing rightward), the neural network learns how to approximate an arbitrarily complex statistical distribution (bottom right) onto a simpler distribution (bottom left) through a series of invertible transforms with tractable Jacobian. In this way, when the training is finished and the application on the testing set starts, one can sample from the simple distribution and recover the more complex posterior distribution by inverting the aforementioned mappings. 

% ***********************
% ****************************************

\section{Posterior coverage tests}\label{app:posterior_coverage}
\begin{figure*}
    \centering
    \includegraphics[width=0.9\textwidth]{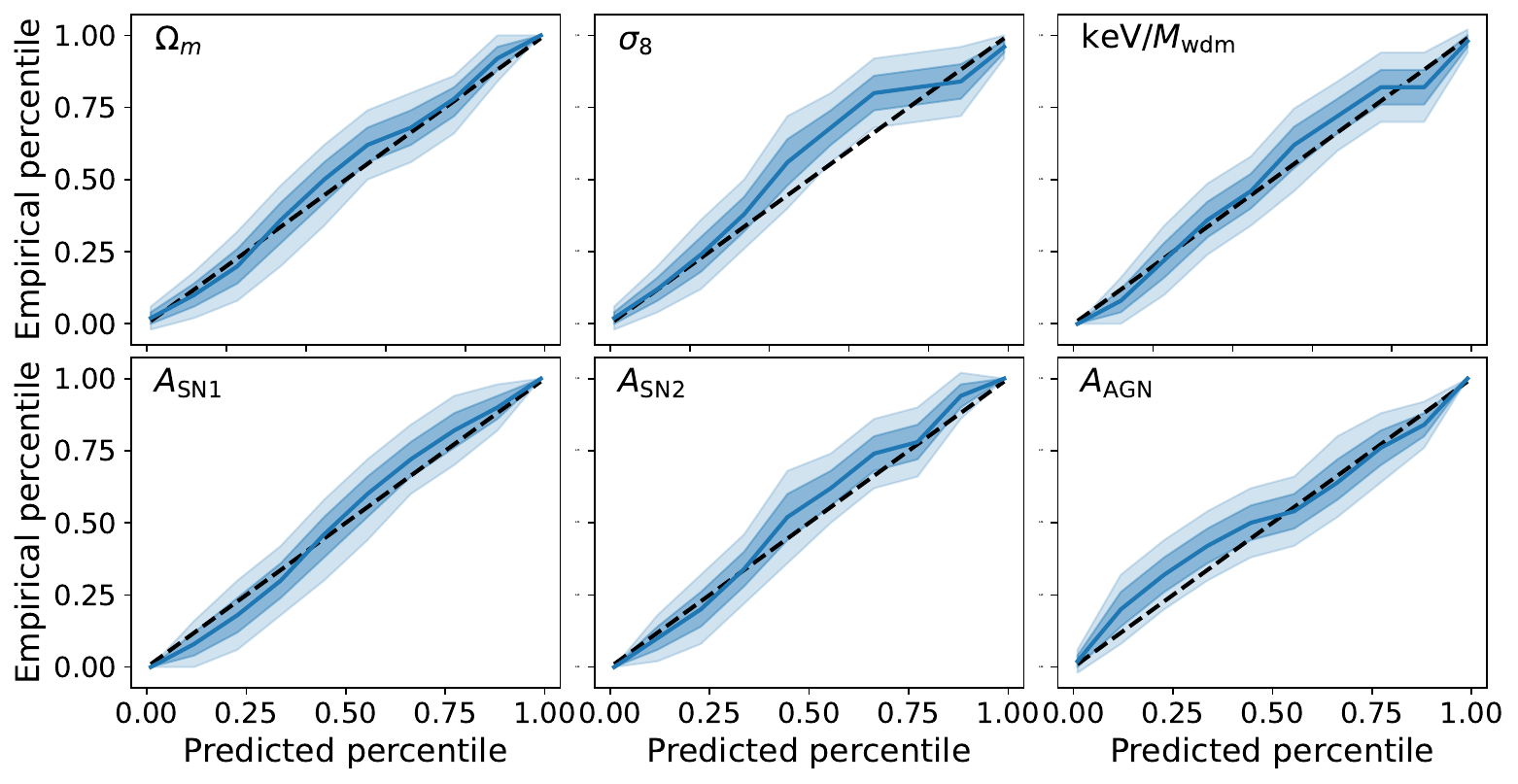}
    \caption{Univariate posterior coverage tests, showing the empirical versus predicted percentiles for the $\Omega_m$ (top left), $\sigma_8$ (top mid), ${\rm keV}/M_{\rm wdm}$ (top right), $A_{\rm SN1}$ (bottom left), $A_{\rm SN2}$ (bottom mid), and $a_{\rm AGN}$ (bottom right) posteriors.}
    \label{fig:coverage}
\end{figure*}

\begin{figure}
    \centering
    \includegraphics[width=0.8\linewidth]{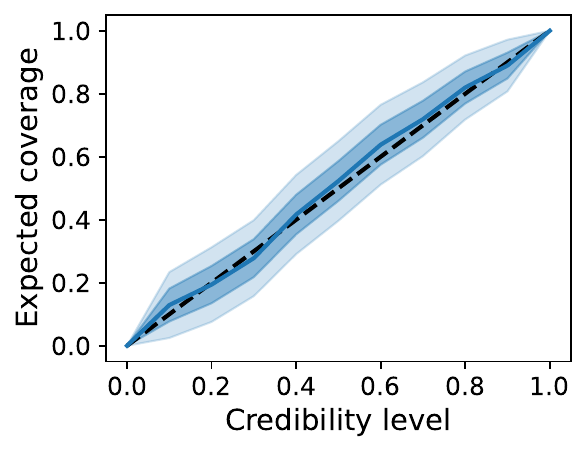}
    \caption{Expected coverage versus credibility level results from TARP.}
    \label{fig:tarp}
\end{figure}

In this Appendix, we address the robustness of the derived posteriors by looking at the posterior coverage. In particular, we check the univariate posterior coverage per each parameter \citep[e.g.,][]{Miller2021,Deistler2022,Hermans2022}, and the combined multivariate `Test of Accuracy with Random Points' \citep[hereafter TARP,][]{Lemos2023}. Figure~\ref{fig:coverage} shows the posterior coverage for $\Omega_m$ (top left), $\sigma_8$ (top mid), ${\rm keV}/M_{\rm wdm}$ (top right), $A_{\rm SN1}$ (bottom left), $A_{\rm SN2}$ (bottom mid), and $A_{\rm AGN}$ (bottom right). The resulting curves show that all the univariate posteriors are well-calibrated. The same results are confirmed also by the TARP results, shown in Figure~\ref{fig:tarp}.

% ****************************************

\section{Observational systematics}\label{app:obs}

In real-data applications, it is crucial to properly account for observational effects and systematics that can affect the measurement of the BTFR, such as:
\begin{itemize}
    \item realistic observational noise: for the measurement of the rotational velocity, this implies mimicking the uncertainty associated with fitting the rotation curve from HI radio data. This can be treated in a simpler way by just statistically assigning uncertainty sampled from distributions , or by forward-modelling the observational radio-astronomical setup and allowing to account for observational systematics more in details
    \item uncertainty on the galaxy inclination, and consequently, to $V_{\rm max}$. In this sense, it is desirable to mimic in the training data set the distribution of inclination as i the data, as well as apply the same inclination quality cuts;
    \item sample selection and the impact of the related systematics on the BFTR; 
    \item for a flux-limited sample, incompleteness effects towards low mass galaxies;
    \item for a volume-limited sample, the effect of cosmic variance. Cosmic variance is already accounted for in \texttt{DREAMS} since different realization were run with different initial random seeds, but it should be properly rescaled for the actual probed effective volume.
\end{itemize}

% ****************************************
% ****************************************
% ****************************************

\section{Detectability of the BTFR from the SKA and its pathfinders}\label{app:ska_detectability}

In this Appendix, we discuss the detectability of the BTFR from the SKA and its precursors. The BTFR is only measurable when the information about both the stellar and the cold gas component is available, i.e. in the presence of commensal optical and H~{\small I} surveys. Given the intrinsic faintness of the 21-cm emission line, from which the H~{\small I} mass can be directly measured, the detectability requirements are mainly driven by the detectability of the 21cm line signal, as the related signal-to-noise ratio is much lower that the one typically yielded by optical photometry at fixed redshift. Current ongoing deep H~{\small I} blind surveys --- such as MIGHTEE-H~{\small I} \citep{Maddox2021} and LADUMA \citep{Blyth2018} --- promise to extend the BTFR detection towards higher redshift. In particular, LADUMA will provide 21-cm line data up to $z\sim 1.4$, and combining a spectral resolution $\Delta v\sim 5.5~{\rm km~s^{-1}}$ with a sensitivity of $\sigma_{\rm RMS}\sim 10~{\mu{\rm Jy~beam^{-1}}}$ achieved through $\sim 3000$ hours integration time, will likely deliver a robust BTFR measurement beyond the local Universe. Future SKA surveys will improve significantly on these figures. Assuming a H~{\small I} survey in SKA-Mid configuration, covering $5,000~{\rm deg}^2$ in $10,000$ hours integration time and with a spectral resolution $\Delta v\sim 2.84(1+z)~{\rm km~s^{-1}}$, one expects millions of H~{\small I} detections at $z\lesssim 0.5$, and therefore the BTFR will be likely to be measured robustly up to $z\sim 0.5$ \citep[see e.g.,][]{Mayor2026}. At $z\gtrsim 0.5$ the forecasts vary considerably on the assumed model, given the large uncertainty on the redshift evolution of the H~{\small I} cosmic density parameter $\Omega_{\rm HI}$. In any case, it will be possible to perform a stacking analysis to detect the BTFR \citep{Meyer2016} up to $z\sim 1.$, which motivates the redshift range probed in this paper. 
% ****************************************

\end{document}